 \documentstyle[aaspptwo,epsf]{article}      
\tighten           
\eqsecnum          
                    
\received{1996}               
\accepted{1996}
\journalid{date }{}       

\def\kms{{\rm{km~s$^{-1}$}\/}}
\def\etal{{\it et al.\/,~}}

\def\ie{{\it i.e.\/,~}}
\def\eg{{\it e.g.\/,~}}
\def\arcdeg{\hbox{$^\circ$}}

\begin{document}    
                    
\title{Diffusive Shock Acceleration in Oblique MHD Shocks: 
Comparison with Monte Carlo Methods and Observations}
                    
\author{Hyesung Kang}
\affil{Department of Earth Sciences, Pusan National University, 
    Pusan 609-735, Korea; e-mail: kang@astrophys.es.pusan.ac.kr\\
\and
\author{T. W. Jones }                            
\affil{Department of Astronomy, University of Minnesota,                
    Minneapolis, MN 55455; e-mail: twj@astro.spa.umn.edu}
    \bigskip}

\begin{abstract}  
We report simulations of diffusive particle acceleration in oblique
magnetohydrodynamical (MHD) shocks. These calculations are based on
extension to oblique shocks of a numerical model for ``thermal leakage''
injection of particles at low energy into the cosmic-ray population. 
That technique, incorporated into a fully dynamical diffusion-convection
formalism, was recently introduced
for parallel shocks by Kang \& Jones (1995).
Here, we have compared results of time dependent numerical simulations
using our technique with Monte Carlo simulations by Ellison, Baring \& Jones 1995
and with {\it in situ} observations from the Ulysses spacecraft of oblique 
interplanetary shocks discussed by Baring \etal (1995). 
Through the success of these comparisons we have demonstrated that our 
{diffusion-convection} method and injection techniques provide a practical tool 
to capture essential physics of the injection process and 
particle acceleration at oblique MHD shocks.

In addition to the diffusion-convection simulations, we have included
time dependent two-fluid simulations for a couple of the shocks to 
demonstrate the basic validity of that formalism in the oblique shock context. 
Using simple models for the two-fluid closure parameters based on test-particle 
considerations, we find good agreement with the dynamical properties of the
more detailed diffusion-convection results. We emphasize, however, that
such two-fluid results can be sensitive to the properties of these closure
parameters when the flows are not truly steady. Furthermore, we 
emphasize through example how the validity of the
two-fluid formalism does not necessarily mean that 
{\it steady-state} two-fluid models
provide a reliable tool for predicting the efficiency of particle
acceleration in real shocks. 

\end{abstract}      
                    

\section{Introduction}                  

The theory of diffusive particle acceleration at collisionless shocks 
has been quite successful in explaining many aspects 
of the cosmic ray (CR) population. 
These include, for example, the nearly power-law spectrum of the CRs 
detected at the top of the atmosphere, the relation
between the break in the power-law around the $\sim 10^{14}$ eV knee 
energy to the maximum
energy of the CRs achievable in supernova remnants (SNRs), and, also
the non-thermal, power-law electron populations deduced from the 
radio synchrotron observations of SNRs (Drury 1983; Blandford \& Eichler 1987;
Berezhko \& Krymskii 1988).
Although the primitive forms of the theory are very straightforward and
robust, the microphysics is actually complex, and there are potentially
important simplifying assumptions built into the various versions of the
theory.  In practice nonlinear interactions between the thermal plasma and the
nonthermal (CR) plasma are also often likely to be essential.
Furthermore, there are certain key features, such as the processes that
inject particles from the thermal plasma into the nonthermal plasma 
(hereafter, simply ``injection'') that are not well understood. 

Over the past several years significant strides have been made in direct
observational tests of diffusive acceleration theory and 
in comparisons between theoretical
models that are sometimes based on very different approaches.
For parallel shocks, in which the ambient magnetic field is aligned with
the shock normal, the applicability of diffusive shock theory is now fairly
well established.
Ellison and collaborators have demonstrated good agreement between
Monte Carlo particle shock simulations and measurements at the earth's 
bow shock (Ellison, M\"obius \& Paschmann 1990),
as well as between Monte Carlo and hybrid plasma shock simulation
techniques (Ellison, \etal 1993).
Recently Kang \& Jones (1995, Paper I) demonstrated that ``continuum transport''
approaches based on the diffusion-convection equation also provide
good models for the same bow shock measurements and that in parallel shocks such
continuum models agree well with both types of particle approaches. 
Paper I also demonstrated that the time-dependent two-fluid derivative of the
diffusion-convection model works well as a dynamical model
for these shocks, as long as the necessary closure parameters are
properly defined.

Paper I, in addition, contained an important step for developing a 
{physically-based} injection model within the diffusive transport formalism.
Particle methods have demonstrated injection to be an integral part of
collisionless shock formation (\eg Jones \& Ellison 1991 and references
therein). 
Continuum models for diffusive acceleration, on the other hand,
have generally depended for practical reasons on an effective separation
of the particle population into distinct thermal and CR components
with the former treated by fluid mechanical techniques and the latter
by approximate plasma kinetic equation techniques or the energy moment of 
that equation (\ie the two-fluid model). 
Paper I introduced a hybrid model of the standard continuum approach. 
It creates a virtual ``injection pool'' of particles that are neither 
fully thermal nor fully ``CR'', but represent the particles leaking
out of the thermal population into the CR component. 
We demonstrated there that this continuum 
``thermal leakage'' injection model could
produce good agreement with simulations done using particle techniques 
and with direct bow shock measurements.
We focus our discussion, by the way, on the ionic particle population,
rather than electrons,
since ions carry most of the momentum flux and seem to capture
the greater share of energy in the shock transition, especially among
the suprathermal, high energy population.
To simplify discussion, we deal only with protons, although the
methods being used can also be applied to other ions.

The situation regarding oblique shocks is not so well
studied and technically more difficult to model. 
Quasi-parallel and quasi-perpendicular collisionless shock 
structures may be quite different (\eg \cite{kenedh85}). 
There are complications in the details of diffusive particle 
propagation (such as anisotropic diffusion with respect to the
magnetic field) and acceleration that derive from the
magnetic field geometry (\eg \cite{jok87}). 
In addition, the magnetic field can apparently
modify and reduce in nonlinear ways the amount of energy that
CRs are able to extract from the flow through highly oblique shocks
(\eg Webb, Drury \& V\"olk 1986; Jun, Clarke
\& Norman 1994; Frank, Jones \& Ryu 1994, 1995).
As a significant step towards a better understanding of the acceleration physics
in oblique shocks Baring, \etal 1995 (BOEF95 hereafter)
have extended the Monte Carlo studies to oblique shocks 
through test-particle simulations in which
a gyro-orbit computation was adopted and large-angle scatterings
were assumed for the particles.
They demonstrated that Monte Carlo shock models with reasonable scattering 
properties can match {\it in situ} observations of
oblique interplanetary shocks from the Ulysses spacecraft.
Ellison, Baring, \& Jones 1995 (EBJ95 hereafter) have applied the same techniques
to study the acceleration rates and injection efficiencies 
in oblique shocks.
They confirmed earlier findings from different methods that, for anisotropic
diffusion, the acceleration rate for individual particles increases 
with the magnetic field obliquity (\eg Jokipii 1987; Frank, Jones \& Ryu 1995).
Thus, {quasi-perpendicular} shocks may be capable of generating higher
energy particles than {quasi-parallel} shocks in a given time, even though their net 
energy extraction efficiencies may be reduced by a strong field. 
In their Monte-Carlo simulations EBJ95 also saw that the
efficiency of ``thermal-leakage''
injection decreases with the obliquity, making it harder for
{quasi-perpendicular} shocks to generate seed CR particles.
They found that the injection rate depends upon the Mach
number, field obliquity angle, and strength of Alfv\'en turbulence 
responsible for scattering.
The last of these enters, because it controls the amount of
cross-field diffusion, which in {quasi-perpendicular} shocks
becomes necessary for particles to escape into the upstream region
in order to be accelerated.
A recent complementary discussion of the characteristics of electron injection at 
quasi-perpendicular shocks has been provided by Levinson (1996).

The present discussion extends the analysis of Paper I
to oblique shocks. We apply our thermal leakage injection model to
oblique shocks and make a preliminary comparison with the behaviors
reported by EBJ95. 
In addition we turn the model to the same interplanetary
shock measurements presented by BOEF95. 
These results will provide a useful foundation for future
studies of oblique shock physics based on continuum models, and
expand on a preliminary report made earlier (Jones \& Kang 1995).
The plan of the paper
is this. In \S 2 we summarize the comparison between particle and
continuum models and how that is important to understanding injection.
Section 3 outlines our numerical methods, while \S 4 presents our
results, and \S 5 provides a summary and conclusion.

\section{Continuum Models and Particle Injection}

In continuum treatments of diffusive shock acceleration, 
the diffusion-convection equation (equation \ref{diffcon}; to be discussed
in \S 3) for the particle momentum distribution is solved along with
the dynamic equations for the underlying plasma. 
They have some distinct practical advantages over Monte Carlo 
methods, especially in time dependent problems and those that involve
complex, or multidimensional flows.  
There are well developed, robust and relatively
inexpensive continuum computational techniques available that can be 
applied very flexibly, for example.
There is a growing literature based on time dependent diffusion-convection
equation treatments of quasi-parallel shocks
(\eg Falle \& Giddings 1987; Kang \& Jones 1991; Kang, Jones \& Ryu 1992;
Duffy, Drury \& V\"olk 1994). 
The simpler and much more economical, two-fluid derivative of the 
diffusion-convection method has seen even greater
application in time dependent dynamical problems, 
because it is practical to use in many complicated situations 
where suitable numerical gasdynamics methods are suitable
(\eg Drury \& Falle 1986; Dorfi 1990, 1991; Jones \& Kang 1990, 1992, 1993; 
Ryu, Kang \& Jones 1993; Jones 1993; Jones, Kang \& Tregillis 1994).
Recently, both of these continuum techniques have been extended to time
dependent MHD and oblique shocks (Frank, Jones \& Ryu 1994, 1995).
Jun, Clarke \& Norman (1994) also reported two-fluid results for 
perpendicular MHD shocks.
Because of the wide-spread use of two-fluid methods their validity
and limits of applicability take on a particular importance. 
Paper I as well as 
some other earlier papers (\eg Duffy, Drury \& V\"olk 1994) have
demonstrated basic validity of the model in quasi-parallel shocks.
There are some important caveats, however, as we shall discuss later.

The diffusion-convection equation is based on the assumption that
a particle momentum distribution is kept almost isotropic in the
local fluid frame by scattering.
When particle velocities are
much greater than the bulk speed of the plasma motion and scattering is
efficient these requirements are very reasonable. 
It is reassuring, but perhaps not terribly surprising that 
statistical particle approaches like Monte Carlo
and continuum (diffusion-convection) methods agree
with each other and with real data in that case.
In the opposite limit of random particle speeds less
than those associated with bulk motion relative to the shock, 
the theoretical situation is again tractable. 
Since charged particles generally are scattered
more rapidly at small velocities, we can expect them to be ``thermalized'' 
effectively and isotropized with respect to
a well-defined mean motion. Then continuum fluid
dynamical computational methods should work well, and statistical
methods should be convergent with them. 
The situation is much more problematic
at ``intermediate'' velocities where particle streaming may be
large, and mean free paths are too long to allow scattering to relax the
distribution to a thermal form. 
(Particles sample a range of environments,
so that no simple thermal equilibrium is appropriate.)
This is the arena of injection, since it includes particles
that are capable of becoming CRs through multiple shock crossings.
The reality of injection in shocks is not much in doubt (\eg Jones \& Ellison 1991),
but despite recent theoretical strides (\eg Malkov \& V\"olk 1995),
the details of injection remain beyond straightforward models.
This problem is difficult because nonlinear interactions between particles,
resonant hydromagnetic waves and the underlying plasma associated with the shock 
formation process itself are very complex and yet to be deciphered fully.
In that context we seek now only a simple, functional model, but one
that captures essential physics of the process.
Previous injection schemes within the continuum formalism have generally 
been based on {\it ad hoc} assumptions that a fixed fraction of the kinetic 
energy flux or the total particle flux through the shock
are transferred from the ``thermal'' to the ``nonthermal'' populations,
so our new approach represents a clear step towards reality.
There is considerable value in developing a serviceable, but physically
based model for injection within the continuum transport paradigm.

The thermal leakage injection model introduced in Paper I is
conceptually simple and represents only a small change from
previous diffusion-convection methods. As before
we simultaneously solve the coupled diffusion-convection/MHD equations.
However, in this new technique we follow the entire proton momentum 
distribution with the diffusion-convection equation, but continuously 
redistribute the  particles at low momenta into a thermal distribution 
according to the pressure and density solution from the MHD equations.
That introduces a population of diffusive particles
at intermediate momenta between the thermal particles and those properly 
termed CR particles.  Since they are diffusive
those particles can sample the fluid
velocity on both sides of the shock, if they are given scattering
properties suitably matched to the numerical shock thickness.
Those intermediate momentum particles
gain energy in a manner that resembles what happens to  CRs, but
their distribution is directly matched onto the thermal distribution,
as it physically must be.
The injection efficiency is determined by the momentum at which this
``injection pool'' distribution is matched to the thermal distribution.
Eventually, we hope to be able to provide an independent model for
this matching, but for now, it can be sufficient to show that
the model successfully reproduces important physical behaviors for
physically reasonable matching conditions.

\section{Model Description}

\subsection{Numerical Methods}

We follow evolution of the particle distribution with the standard
diffusion-convection equation (\eg Parker 1965; \cite{ski75}; \cite{jok}),
\begin{equation}
{df\over dt} = {1\over3}  \vec\nabla\cdot (\vec u~+~\vec u_w)  p {\partial f\over
\partial p} + \vec\nabla\cdot (\kappa \vec\nabla f) - 
\vec u_w\cdot\nabla f,
\label{diffcon}
\end{equation} 
where $f(x,p,t)$ is the isotropic part of the distribution function
measured in
the convected frame, $\vec u + \vec u_w$, while $d/dt$ is the total 
time derivative in the fluid frame, $\vec u$. The propagation of scattering
centers relative to the plasma is represented by $\vec u_w$. Generally,
the scattering centers are assumed to be Alfv\'en waves resonant
with the particles, so $\vec u_w$ represents the center-of-momentum motion 
of those Alfv\'en waves.
Our simulations assume planar symmetry, so $\kappa = \kappa(p)$, termed
the diffusion coefficient, is the projection
of the spatial diffusion tensor onto the shock normal.
That direction is taken to be parallel to the $x$ axis.
Throughout the paper we express momentum, $p$,
in units of the proton rest mass energy, $mc^2/c~(=9.38\times 10^5 {\rm keV}/c$),
and the distribution function $f$ in units of the 
particle number density, so that $4\pi \int f~p^2~dp = \rho/m$.
We solve equation \ref{diffcon} on an Eulerian grid using a second-order 
combined Lagrangian 
Crank-Nicholson/monotone-remap scheme whose details can be found 
in Kang \& Jones (1991) and Frank \etal (1995), respectively.
The distribution $f(p)$ is supposed to exist over a range
$p_0 \le p \le p_3$ that includes both the thermal distribution
and the CR distribution.
The thermal distribution expressed in terms of $g(p) = p^4 f(p)$, 
which measures the partial pressure, $dP/dp$, has its maximum at
$p_{th} = \sqrt{4\tilde T} = \sqrt{4m~k_B~T}/mc$, 
where the gas temperature $\tilde T$ is expressed in units $mc^2/k_B$. 
We will identify those particles dynamically as CRs that
satisfy $p\ge p_2 >> p_{th}$, and will establish $p_2$ below. 

The dynamics of the underlying plasma is followed by an explicit, 
second-order accurate MHD code based on a conservative up-winded,
Total Variation Diminishing (TVD) scheme (Ryu \& Jones 1995) that has been
modified to include the dynamic effects of the CR pressure
(Frank \etal 1995). 
Readers are referred to their papers for the basic MHD equations
and the detailed description of the numerical method.
Based on a linear Riemann solver used to compute ``up-winded'' mass, momentum and
energy fluxes at zone boundaries, the code generally captures cleanly all 
the families of MHD discontinuities.
Strong shocks are usually contained within 2 to 3 zones, and other 
discontinuities within a slightly broader space 
($\sim 4-10$ zones, depending on the feature).
The code is conservative in the sense that it maintains exact net fluxes 
through the grid to machine accuracy. The TVD label refers to the manner
in which the code avoids introducing physically spurious oscillations
by preserving monotonicity in physical flow variables through discontinuities.

There is one addition to the code discussed in Frank \etal (1995);
namely, ``Alfv\'en Wave Transport'' (AWT) terms, 
as represented in equation \ref{diffcon} by $\vec u_w$.
Those are handled in the same way as discussed by Jones (1993) and Paper I 
for parallel shocks, with the
proviso that $\vec u_w$ aligns with the local magnetic field vector.
Additional AWT terms provide for gas heating due to dissipation of
the energy transferred {\it from} CRs (the last term in equation \ref{diffcon})
to Alfv\'en waves (thence to the plasma) 
and also transport of the energy and momentum
content within the waves. In the present simulations we have neglected
the energy and momentum carried explicitly within the wave field, but have
included the energy and momentum passed {\it through} the wave field
(see Jones 1993 for details).
The magnetic field in this problem lies within a single plane containing
the shock normal direction, $\hat x$. So, without loss of generality
we can define the magnetic field to be within the $x-z$ plane.

For oblique MHD shocks the diffusion coefficient takes the 
standard form
\begin{equation}
\label{kappa}
 \kappa = \kappa_{\parallel} \cos^2 \theta + \kappa_{\perp} \sin^2 \theta,
\end{equation}
where $\parallel$ and $\perp$ refer to diffusion along and across the
magnetic field direction, respectively, and $\theta= \arctan ( B_z/B_x)$.
Following Jokipii (1987),
we assume a parallel diffusion coefficient of the form
$\kappa_{\parallel}={1\over 3}\lambda_{\parallel} v $, with the
scattering length, $\lambda_{\parallel} = N r_g$, where
$r_g$ is the gyro-radius of a particle and $v$ is its speed. 
Then, from standard kinetic theory,
the ratio of the parallel to perpendicular components is determined
by the ratio $N>1$ as
\begin{equation}
\kappa_{\perp}/ \kappa_{\parallel} =
 ~[1+ (\lambda_{\parallel}/ r_g)^2]^{-1} = (1+N^2)^{-1}. 
\end{equation}
Equation \ref{kappa} can be rewritten as 
\begin{equation}
 \kappa =[ N \cos^2 \theta + ({N \over{1+N^2}}) \sin^2 \theta] \kappa_B,
\end{equation}
where $\kappa_B = \frac{1}{3} r_g v$ is the Bohm diffusion coefficient.
The limit $N \to 0$ corresponds to Bohm diffusion, where
$\kappa_{\perp} \sim \kappa_{\parallel}$.
Cross-field diffusion is determined in this model by the strength of
Alfv\'enic turbulence, since,
$N \sim E_B/(kE_{wk})$, where $E_B$ and $E_{wk}$ are
the total energy density in magnetic fields and the Alfv\'en wave
energy density at the resonant wave number, $k$, respectively. 
When scattering is weak, so that $N>>1$, there is little cross-field
diffusion, whereas strong scattering leads to cross field diffusion
comparable to field-aligned diffusion.
If $N$ is a constant, it follows for nonrelativistic particles that
$\kappa \propto p^2$.

\subsection{A Numerical Injection Model}

As we stressed above, the detailed physics of the injection
process is not yet well understood, and diffusive transport models
cannot, by themselves, accurately treat the particles directly involved in the process.
So, as a practical approximation we assume a simple but reasonable scenario 
in which a small population of near-thermal particles gain excess energy via 
interactions with resonant waves and form a 
suprathermal tail on the Maxwellian distribution in the vicinity of
the shock front.
They provide the seed particles injected into the CR population.
As noted before, in our model 
the particle distribution over the full range of momenta including
the thermal plasma is followed explicitly. 
Below a certain momentum (in units of $mc$),
$p_1=c_1~p_{th} = c_1~\sqrt(4\tilde T)=c_1 \sqrt{4 m k_B T }/mc$,
chosen high enough to include most of the postshock thermal
population, the distribution is forced to maintain a Maxwellian form
consistent with the local gas density and pressure determined from the
MHD equations. 
Above $p_1$ particles are allowed to evolve according to the 
diffusion-convection equation, while only for
$p\ge p_2 > p_1$ are they considered dynamically as CRs.
Particles between $p_1$ and $p_2$, thus, constitute ``candidate'' 
CR particles, because they are not locally thermalized. 
They can be injected into the CR population
by crossing a momentum boundary
at $p=p_2$ through flow compression. 
Below $p_1$ particles are compressed adiabatically (reversibly) by the flow,
except within the shock discontinuity, where the shock jump conditions 
demand that the compression be irreversible. 
Above $p_1$, on the other hand, compression leads to
irreversible energy changes in the particles, because diffusion is
irreversible. 
This combination of effects is the source of energy for the particle acceleration, of course.

We emphasize that the distribution function for thermal particles is used 
only to provide the reference population needed to match onto the intermediate population,
while the thermal pressure, $P_g$, is 
included and handled through the MHD equations.
On the other hand, the intermediate population only provides the seed particles
for CR particles and has no dynamical effects on the flow in our method.
Although the particle distribution is continuous over the full range of momenta,
in continuum treatments short of a full solution to the Boltzmann equation
one needs to separate the pressure due to thermal
particles ($P_g$) and that due to CR particles ($P_c$), because their dynamical
behaviors are different.
This, of course, necessitates a definition for the CR population. We have
done that by choosing $p_2$ as the arbitrary boundary.
At early stages of acceleration, when the particle distribution is
almost Maxwellian, the small CR pressure is sensitively 
dependent upon the chosen value of $p_2$, but $P_c$ is dynamically insignificant
then.  On the other hand, When $P_c$ becomes large enough to be important, it becomes much
less dependent upon $p_2$, because particles of much higher energy dominate
the CR pressure.
As a result, our calculations are not critically dependent upon the parameter
$p_2$. So, it is most convenient numerically
to fix the value of $p_2$ $\sim (3-4) p_{th,i}$, where $p_{th,i}$ is
the thermal peak momentum of the postshock gas of the initial pure 
gasdynamic shock.
The particle supply in the intermediate momentum pool is sensitively 
controlled by the parameter $p_1$, since $f(p_1)$ is part of the
exponential tail of the postshock Maxwellian distribution.
According to comparison tests with measurements of a
parallel occurrence of the earth's bow shock and with shocks
computed by ``particle'' methods (see Paper I), appropriate values
of the related scaling parameter, $c_1=p_1/p_{th}$, fall in the very 
reasonable range $c_1 = 1.5-2$.  
Thus the value of $p_1$ varies in time and space along with the local gas 
temperature. 

The model further requires us to match the numerical shock thickness, $\delta x$,
to particle scattering properties, since the numerically realized injection 
rate will depend upon the ratio $\lambda(p_1)/\delta x$
(see Paper I).
Above $p_1$ particles are formally diffusive, but unless the
scattering lengths of these particles projected onto the shock
normal, $\lambda(p)\cos{\theta}$, exceed 
$\delta x$, they cannot be effectively accelerated by the 
Fermi process.
On the other hand thermal particles should not be able to cross
the shock within a projected scattering length, since they should then
not form into a Maxwellian distribution.
Hence, the numerical shock must be thicker than the projected scattering length
of thermal particles, but thinner than the projected scattering length of CRs.
The structure and thickness of {\it real} shocks will be dependent 
upon the details of the strength and geometry of the field, 
degree of turbulence, the strength of the shock, for example. 
That issue is beyond the scope of this study.
The specifics adopted for the numerical shock thickness will be
given later for each case.
This ``thermal leakage'' type injection model is rather simple,
but, according to the results reported in Paper I, apparently able 
to capture essential 
characteristics of real injection processes, provided that one 
makes a reasonable choice for the free parameter $p_1$.

\subsection{Initial and Boundary Conditions}

For our simulations the initial flow is specified by a simple discontinuous 
MHD shock using standard jump
conditions, which can be found from MHD Riemann solutions, for example,
(see Ryu \& Jones 1995). 
The shock faces to the right, so that velocities
along the $x$ axis are negative when the shock is nearly at rest
in the grid. All of our shocks start at rest in the grid, but
those developing dynamically significant CR pressure become temporarily
``over-compressed'', as expected,
causing them to drift slowly to the left. Three fluid parameters
are needed to define the shocks. Those can be the sonic Mach number, 
$M_1 = u_{1x}/c_{s1}$, the strength of the  upstream magnetic field, 
$B_1$ and the upstream obliquity of the magnetic field, $
\theta_1 = \arctan{B_{1x}/B_{1z}}$.
The sound speed is $c_{s1} = \sqrt{\gamma P_g/\rho}$, where $P_g$ is
the gas pressure, $\rho$ is the gas density and $\gamma$ is the
gas adiabatic index, taken to be $\frac{5}{3}$.
We also assume that initially the particle distribution function, $f(p)$,
is Maxwellian everywhere, with a temperature, $\tilde T =(P_g m)/(\rho k_B)$. 
Thus, there are no CRs initially. They are injected through
thermal leakage as part of the process of shock evolution. 
The above definition of the temperature, which was used in both EBJ95 and
BOEF95 implies that the electron pressure is negligible 
compared to the proton pressure. More recently, Baring \etal 1996 (BOEF96)
have recomputed the properties of the Ulysses-observed shocks, including
finite electron pressure. Although that has changed some of the
shock parameters, it should not alter any of our conclusions. 
The influence of a finite electron pressure is certainly straightforward
to include when warranted.

The MHD variables are assumed to be continuous across the left and 
right boundaries of the spatial grid.
This is a good assumption, since the shock is approximately at rest
in the middle of the grid, keeping any gradients in flow variables small
near the boundaries.
The particle distribution, $f(x,p)$, is also assumed to be continuous across 
the boundaries, which means diffusive particle fluxes vanish there.
This no-flux boundary condition is
numerically simple and robust for the diffusion-convection equation.
It remains reasonable as long as the particles are confined near the
shock and away from the boundaries. 

The boundary condition for $f(p)$ just below $p_0$ is not relevant 
here, since the distribution is redefined continuously by the 
Maxwellian function at each time step for $p<p_1$.
At the highest momentum boundary, we assume $f(p)=f(p_3)$ for $p>p_3$.
This condition is not very crucial either, since the divergence
of the flow is rare around the shocks in plane-parallel geometry
considered here.  

\section{Results}

\subsection{Comparison with Monte Carlo Simulations}

EBJ95 have calculated, by test-particle Monte Carlo simulations, the 
efficiency of injection at oblique shocks as a function
of Mach number, $M_1$; field obliquity, $\theta_1$; and the degree of
cross-field diffusion (as measured by $N= \lambda/r_g$). 
They found the injection to be more efficient for lower 
Mach numbers, for smaller obliquities and for stronger cross-field 
diffusion (\ie smaller $N$).
In their Fig. 5 they showed the downstream integral density distribution for
particles accelerated in strong shocks for a range of obliquity.   
This information can be compared directly with
the particle distribution functions of our simulations.
Thus, we chose these shocks as the comparison models and found the 
value of $c_1$ for each value of $\theta_1$ that gives the
best fit to their results. 
The common shock parameters are 
$u_{1x} = 500$ km~s$^{-1}$, $M_1=100$, $N=\lambda/r_g=100$, 
and $B_1=10^{-8}$ Gauss.
This represents a very strong shock in the limit of weak cross-field
diffusion and weak magnetic field.
The obliquity values considered are 
$\theta_1=0$\arcdeg, 20\arcdeg, 30\arcdeg, and 35\arcdeg; so that all are
``quasi-parallel'' shocks.
Larger obliquities were not considered by EBJ95 for this shock
system, since CR injection was found to be completely suppressed.

The EBJ95 simulations were test-particle, so for this
comparison test only, we turned off the dynamic evolution of the flow 
and kept the shock structure as the
initial discontinuous jump (thus, with no CR pressure feedback, 
even though our code is designed to include fully the dynamical 
contributions of the CR).
For these test-particle runs the shock thickness is effectively
one grid cell. 
Since the shock thickness should be of order the mean scattering
length of the postshock thermal momentum, $p_{th}$, we adjust the grid
spacing to be this length (\eg $\Delta x = 
\lambda(p_{th})= N r_g(p_{th})$).
All physical lengths in this problem scale with $\lambda(p_{th})$,
so this model for the shock thickness will make the injection process
scale with $N$.
Our comparisons with EBJ95 were with Monte Carlo simulations that did not include 
AWT, so we turned those effects off for this particular set of
continuum transport simulations.

Monte Carlo simulations intrinsically consider a steady state, while
our calculations are time-dependent. 
Fig. 5 in EBJ95 shows the Monte Carlo, integral density distribution up to 1000 keV.
In order to make a good comparison
we should integrate our simulations for a time 
comparable to that required to accelerate a thermal 
particle to $E > 1000$ keV. 
In practice, however, these become fairly expensive for cases with smaller obliquity,
$\theta_1$, because the integration time is longer ($t_{acc} \propto \kappa$),
and so a greater spatial length is required to keep the CR particle distribution small
at the boundaries.
Thus we evolved each shock for a time needed to accelerate particles
to $E\sim {\rm~a~few}~\times~100$ keV. That corresponds to
$(t/10^8{\rm s})=$ 12, 8, 6, and 4  for
$\theta_1=0$\arcdeg, 20\arcdeg, 30\arcdeg, and 35\arcdeg, respectively.
Fig. 1 shows the resulting particle distributions at the shock position
for the times and values of $\theta_1$ specified above.
The cases shown here are the models with $c_1$ chosen to match the results
of EBJ95. 
The top panel shows $p^4 f(x_s,p)$, where $x_s$ is the shock position. 
The canonical test-particle distribution is a power-law; in this strong 
shock case, $f(p)\propto p^4$.
The distributions shown in Fig. 1, however, are somewhat steeper than
this canonical power-law, because the computed momentum range is
finite and because the computed time interval does not a real
steady state to be achieved.
The slopes of pow-law fits of these distributions at 
$p \sim 1.5\times10^{-2}$, for example, are $q \sim  4.06 - 4.1$, 
where $f(p)\propto p^{-q}$. 
Following EBJ95, the bottom panel shows
the integral of the distribution function above a given kinetic energy, 
expressed in units of keV; namely,
$n(>E) = 4\pi \int^\infty_p p'^2 f(p')dp'$,
where $E = mc^2 (\sqrt{p^2 + 1} - 1)$.
The bottom panel also includes the analogous results reported by EBJ95 for their
simulations.

As mentioned before, our spectra start to cut off above 
$E\sim {\rm~a~few}~\times~100$ keV, due to limited evolution time,
while the steady-state, EBJ95 Monte Carlo results 
extend to higher kinetic energy values.
More recently Ellison, Baring \& Jones  1996 (EBJ96) have extended their test-particle
simulations to fully dynamical Monte Carlo simulations. In those simulations
they included a ``Free Escape Boundary'' (FEB), which removes particles
that propagate ``too far'' upstream from the shock. That preferentially
removes the highest energy particles, since they have the longest
scattering lengths. The net result is an energy cutoff that 
qualitatively resembles the finite-time cutoff observed in the
distributions we show in Fig. 1.
Notice that the high energy side of the EBJ95 ``quasi-thermal'' distributions
cut off more sharply than Maxwellian. 
This presumably results from the rapidly increasing rate of thermal
particle ``leakage'' with momentum in the Monte Carlo simulations.
Our distribution, on the other hand, is not allowed to deviate from 
the Maxwellian form below $p_1$ corresponding to $E\sim 2$ keV, 
and we simply match the nonthermal distribution to it.
But we see that within an energy factor of 2 or 3 of the thermal
peak ($p_{th} \approx 1.5\times 10^{-3}$, $E_{th} \approx 1$ keV)
our distribution converges fairly well to that found by EBJ95, 
below the cutoff imposed by finite acceleration time.
Thus, on the whole our model shows itself to be a reasonable way to mimic 
the injection and acceleration processes. 
It produces a consistent particle spectrum at energies
higher than thermal energies, in agreement with the Monte Carlo
simulations, even though the details of the
injection of suprathermal particles are not included. 

The values of $c_1$ adopted for the test-particle simulations that
fit best with EBJ95 results are 1.4, 1.65, 2.0, and
2.3 for $\theta_1=0$\arcdeg, 20\arcdeg, 30\arcdeg, and 35\arcdeg, 
respectively.
The increasing values of $c_1$ for higher obliquity are required to
reduce the injection rates for those shocks. From Fig. 1
or from EBJ95 Figs 5 \& 6 it is apparent that the injection rate decreases
by about two orders of magnitude between $\theta_1=0$\arcdeg~and
$\theta_1=35$\arcdeg~for this Mach number and $N$ value. In fact, EBJ95
argue within the test-particle picture that above 
$\theta_1 \sim 30$\arcdeg, injection within
strong shocks may be completely suppressed in the
absence of cross-field diffusion. The reason for the 
obliquity dependence is that particles propagate along field lines
until they scatter, except for a drift along the shock plane that can be eliminated by
referring to the so-called de Hoffmann-Teller frame.
In strong, oblique shocks $\tan{\theta_2} = r\tan{\theta_1}$,
where $\theta_2$ is the downstream field obliquity and $r$ is the shock 
compression ratio; that is, the
downstream obliquity is greater than the upstream obliquity and
downstream particle motions are more nearly along the shock plane. 
Thus, as the obliquity
increases, a relatively larger total particle speed after an
initial scattering
is needed to enable a particle to ``swim upstream'' fast enough to re-cross the 
shock from downstream. This tendency reduces the number of particles
available for injection (Baring, Ellison \& Jones 1994). 
In our simulations the same
effect is established by higher values of $c_1$ for higher obliquity. 
As $c_1$ increases thermal leakage is reduced, because the number of 
particles in the injection pool is reduced. 
~From the above explanation it is clear in this model that injection
is less sensitive to obliquity when the Mach number is smaller
or when the scattering is stronger (that is $N$ is smaller).
EBJ95 found in those situations that the injection rate is also
greater.
This implies that smaller values of $c_1$ should be chosen in our model for 
smaller $M$ and for smaller $N$,  
since the injection efficiency is mostly controlled by the
value of $c_1$.
However, we do not attempt here to find a quantitative dependence of $c_1$
on $M$ or $N$, since the information presented in EBJ95 is insufficient
for that.
Also, the best-fit values of $c_1$ could vary with the numerical
shock thickness. We leave for the future a more detailed analysis of these
model properties.

The above simulations, both ours and those in EBJ95, were of a test-particle
character. On the other hand, it is clear that
the energy represented in the super-thermal particle distributions
is a substantial fraction of the total.
Thus, test-particle results are not very meaningful as a
measure of the properties of real shocks of this kind. 
This is not surprising, since previous studies of strong 
gasdynamic CR shocks have found them to be very efficient at
transferring energy from the flow to CRs (\eg Drury \& V\"olk 1981).

To gain some insights into the properties of these shocks when they are constructed
self-consistently, we repeated our simulations, but with the fully
dynamic version of our MHD/diffusion-convection code. To keep the
tests simple we used the same values of $c_1$ to model injection as in the
test-particle simulations.
For an obliquity less than 20\arcdeg, however, more than 10\% of the 
particle number density is within the CR population (see Fig.1) by the end of the
simulation interval, using the injection rates found by EBJ95.
So, one more assumption of the MHD/diffusion-convection approach is invalid; 
namely, that the inertia within the high energy, CR population can be neglected.
For this reason we have done dynamic test runs only for 
$\theta_1=$30\arcdeg.  For convenience in later discussions we refer to this model
shock as {\it EBJ95-D}.
Unlike the test-particle runs where the shock is one cell thick,
strong shocks in fully dynamic runs are captured within about two cells
in our code.
Thus, for these tests the grid is adjusted so that a cell has thickness,
$\Delta x=0.5 \lambda(p_{th})$  
in order again to match the
numerical shock thickness to the scattering length for thermal particles;
\ie $\delta x \approx \lambda(p_{th})$.

Fig. 2 shows the flow structure around the {\it EBJ95-D} shock at $t/(10^7s)=4,
~8,$ and $12$, along with
the initial MHD shock jump at $t=0$.
The frame of reference is chosen so that the shock is 
at rest without CR modification to the flow.
The physical variables are expressed for simplicity
in units of the following normalization constants: $L_o = 5\times 10^{15}$ cm,
$\rho_o=1.67\times10^{-24} {\rm g~cm^{-3}}$, $u_o = 5\times 10^2 {\rm
km~s^{-1}}$, and $P_{go}=4.175\times10^{-9} {\rm erg~cm^{-3}}$. 
The numerical grid extends from 0.0 to 3.0 in units of $L_o$. Only
the region between 0.0 and 2.0 is shown in the figure, however.
The assumed value of $c_1=2.0$, is the same as for the test-particle
run.
Predictably, the CR pressure is dynamically important for this 
shock, so that it has a clear precursor. 
Similarly, the maximum compression is 4.6 instead of the test-particle 
value of 4 and the postshock gas pressure is lower than that of the
initial shock.
Note at $t/(10^7s)=12$, that the shock structure is still evolving 
rapidly and the postshock CR pressure is already about 30 \% of the 
gas pressure.
Thus, it is clear that the test particle approximation is not valid 
even for this obliquity.
Since the shock structure has been modified by the CR pressure from the
initial shock jump, the shock is moving slowly to the left from the
initial position.
As time goes on, the modified structure extends downstream (to the left 
in Fig. 2) due to advection, while the precursor in the velocity and
the CR pressure extends upstream via the diffusion of highest 
energy particles.

Fig. 3 provides a comparison of the particle distribution and the 
integrated particle
density for {\it EBJ95-D}, as well as the test-particle calculation
shown in Fig. 2 with the same initial conditions. 
We first note that the gas temperature is lower, so the peak momentum of the
Maxwellian distribution, $p_{th}$, is lower in the dynamic run, as we expect
from the lower postshock gas pressure and higher compression 
shown in Fig. 2.
The postshock gas is colder in the dynamic run and so
the particles in the thermal tail have smaller $r_g$, while 
the shock numerical thickness is the same length in both runs.
Thus they are less likely to be able to cross the shock in
the dynamic run than in the test particle run.
This will reduce the injection rate in the fully dynamic run. 
Therefore, we expect for similar reasons that the injection
rates in fully dynamic Monte Carlo simulations would decrease from
those given in the EBJ95 test particle simulations,
especially for small obliquities.

\subsection{Comparison with Ulysses Observations}

BOEF95 have compared proton distributions 
measured directly by the Ulysses spacecraft at oblique interplanetary shocks 
with results from Monte Carlo simulations of similar shocks.
The BOEF95 simulations are also test-particle ones. 
We have adopted the same shock parameters as they obtained,
and calculated the time-dependent evolution of the particle
distribution functions.
Our runs include fully the dynamical feedback of CRs on the shock structure; however,
because we are comparing our results with the Ulysses data rather than the
Monte Carlo simulations.
The resulting CR-induced flow modifications are small enough that we do not 
expect significant differences in the particle distributions from a comparable test-particle simulation.
Similarly, we have included the effects of Alfv\'en wave transport, since it would
presumably be present in the real interplanetary shocks.

BOEF95 have studied two shocks. 
For the first shock, observed on April 7, 1991, (hereafter {\it BOEF95-1})
the following properties are assumed: shock velocity, $V_s=153$ \kms, sonic Mach number, $M_s=6.9$;
Alfv\'enic Mach number, $M_A=3.1$; upstream field strength, $B=30 \mu$G;
upstream particle density, $n_1=1.756{\rm cm}^{-3}$;
upstream ion temperature, $T_1=3.57\times 10^4$K, and
magnetic obliquity, $\theta_1 = 77$\arcdeg.
The second shock, which was observed on April 28, 1991 (hereafter {\it BOEF95-2}) 
was a bit weaker than the first shock.
This shock is initiated with these conditions: shock velocity, $V_s=165$ \kms; 
sonic Mach number, $M_s=3.9$; Alfv\'enic Mach number, $M_A=2.2$; upstream 
field strength, $B=20 \mu$G; upstream particle density, $n_1=0.338{\rm cm}^{-3}$,
upstream ion temperature, $T_1=1.3\times 10^5$K, and
magnetic obliquity, $\theta_1 = 75$\arcdeg.

The grid spacing in all runs for these two shock models is set so that
$\Delta x=1/2 r_g(p_{th})$ for the {\it BOEF95-1} shock, and 
$\Delta x=1/3 r_g(p_{th})$ for {\it BOEF95-2},
independent of the value of $N$.
Since the shock spreads over 3-4 cells in these relatively weak shocks, 
the effective numerical shock thickness, $\delta x \sim (1- 2) r_g(p_{th})$.
These values of $\Delta x$ are necessary to produce
particle fluxes matching the observations.
For $\Delta x$ twice these values, for example, 
particle fluxes are too low to match the observations with any reasonable
choices of $c_1$. 
We note below that the best fits to the Ulysses
data correspond to $N = 4$ for {\it BOEF95-1} and $N = 9$ for {\it BOEF95-2}.
These shocks are quasi-perpendicular, with $\theta_1 \approx 75$\arcdeg
and $\theta_2 \approx 85$\arcdeg, so that as a particle streams a
distance $\lambda$ along a field line, it moves along the shock normal
a distance $\lambda\cos{\theta} \sim N \cos{\theta} r_g$ which is
$\sim (1-2) r_g$.
Thus, it makes sense that the required numerical shock thickness, 
$\delta x \sim (1-2) r_g(p_{th})$.

The value of $p_2$ was fixed at $3 p_{th,i}$, 
where $p_{th,i}$ is the momentum at which the Maxwellian distribution
peaks in the downstream region for the initial shock.
The CR pressure is small compared to the gas pressure, and consequently
not very different from test-particle conditions.
Thus, the choice of $p_2$ substantially affects neither the flow dynamics nor the particle spectrum.
    
In order to transform the particle distribution function (which is
isotropic to lowest order in the local fluid frame) to the particle count 
rate in the space craft frame, we need to know 
the velocity of the downstream flow relative to
the spacecraft. 
That velocity is difficult to compute accurately from the information
available, so we chose it to match the particle velocity,
$V_{peak}=500$\kms at the peak of the Maxwellian distribution.

Fig. 4 shows the computed and measured
omni-directional particle flux in the spacecraft frame divided by 
the particle momentum cubed, $p^3$, and also the computed
particle distribution function for the {\it BOEF95-1} shock. 
The filled dots are the Ulysses data taken from Fig. 1 of BOEF95. 
Our results are shown at $t=6$ minutes. 
For the velocity range $500<V_p<2000$ \kms, 
the simulated particle flux has reached nearly steady values 
from an initial Maxwellian form after 5 minutes.
Three values of $N=\lambda/r_g=4$, 20 and 40 were tried
while keeping $c_1=1.6$. 
For the fourth run, $N=4$ and $c_1=2.0$ were chosen. 
The grid spacing is the same and so the shock thickness is about the same 
for all four cases. 
All except the $N = 40$ run produce acceptable fits to the Ulysses data, 
although the $N = 4$ is somewhat the best. That value of $N$ was also 
preferred by BOEF95 from their Monte Carlo simulations.
The similar comparison for {\it BOEF95-2} shock is presented in Fig. 5.
The same quantities are plotted as in Fig. 4. 
Now our results correspond to a shock evolution time, $t=10$ minutes.
The value of $c_1$ for the best fit is again 1.6. 
Three values of $N$ are compared; namely $N = 9,~20$ and 40. 
For a fourth run $N=9$ and $c_1=2.0$ were used. 
As in BOEF95, the simulated particle fluxes seem to 
agree best with observations when small values of $N$ are used.
Although these calculations include full dynamic effects of CRs,
the modification to the flow structure is insignificant as shown in
Fig. 6.  But the slight reduction in postshock pressure and temperature
in dynamic calculations means somewhat smaller injection rate compared 
to the test-particle simulations.  
For test-particle simulations, the particle flux shown
in Figs. 4-5 could be about 50 \% larger than that of dynamic runs
for the velocity near 1000 \kms, for example.

While, in both examples, the comparisons of each case with the BOEF95 
fluxes are fairly similar, we can see in the $p^4f(p)$ plots that 
smaller $N$ leads to higher momentum particles at a given time. 
That is simply due to the fact that smaller $N$ leads to smaller
$\kappa$ (see equation \ref{kappa}), and consequently a smaller 
acceleration time, since the individual particle acceleration time, 
$t_{acc} \propto \kappa$ (see, \eg \cite{lagces83}). 
The particle flux near and above $p_1$, however,
increases with the values of $N$ for three runs with $c_1=1.6$.
The particles have larger mean free paths for larger $N$ 
and so have higher probability to cross the shock,
since the shock thickness is about the same for all runs. 
This leads to a higher particle flux in the injection pool and so a higher
injection rate. 
However, the sensitivity to $N$ is rather weak in our model, 
compared to that to $c_1$, for a given shock thickness 
and for a given value of $c_1$.
The $c_1 = 2.0$ cases produce fewer CRs, 
but accelerate them to the same momenta as the same $N$ and $c_1 = 1.6$ 
cases.
That is because $c_1 = 2.0$ places the transition from thermal to 
nonthermal particles farther into the Maxwellian tail of the 
postshock distribution, and thus, reduces the population 
of the injection pool.
This shows that the injection rate is mostly controlled
by the choice of $c_1$ for a given shock thickness. 
In our numerical injection model we do not have a self-consistent way to
determine the best value of $c_1$ for a given value of $N$,
while in Monte Carlo simulations the injection is treated self-
consistently.
On the other hand, the fact that the numerical shock thickness 
must be relatively thin to produce consistent fluxes
(\ie $\delta x \sim (1-2) r_g(p_{th})$, so that $N\cos{\theta} \sim 1$
for the best fits with $N$)
could imply that the observed particle flux cannot be explained
if the scattering turbulence is weak (\eg $N \gg 1$).
This is consistent with the conclusions of BOEF95.

Again these comparison calculations
have shown that the diffusion-convection formalism with our new 
injection scheme and a reasonable set of scattering and injection parameters 
can reproduce the particle 
injection and acceleration processes in real oblique MHD shocks.
The detailed dependence of our calculations upon the model parameters
such as $c_1$ and grid spacing should not be overemphasized, since
our model is not intended to represent the detailed microphysics of the
injection and shock formation processes, but rather only to try to
capture the outcomes reasonably well.

\subsection{Two-Fluid Comparisons}

Beginning from the above successes,
it is useful to provide direct comparisons between the diffusion-convection
simulations and the simpler two-fluid versions of them. 
Two-fluid methods have been especially useful in complex time
dependent applications, such as the evolution of supernova remnants
(\eg \cite{dor91};~\cite{jonkan92}). They are currently the only
practical method of calculating multidimensional CR-modified flows
(\eg \cite{jonkant94}).
As mentioned in the introduction, there has been some controversy in 
the past about the conditions under which two-fluid
methods can provide reliable dynamical solutions for diffusive shock structures. 
Paper I addressed some of these issues in the context of parallel shocks,
and identifies some of the background literature. 
We demonstrated there the basic agreement between two-fluid and 
diffusion-convection methods. 
Arguments are sometimes expressed that momentum-dependent, cross-field 
diffusion in oblique shocks could invalidate the fluid-like CR behaviors 
implicit in the two-fluid formalism.
To the best of our knowledge the only previous comparisons of the methods for oblique
MHD shocks were by Frank, Jones and Ryu (1995). 
They considered only a case with a momentum independent diffusion 
coefficient and one with weak momentum dependence, $\kappa \propto p^{1/2}$.
Thus, we provide here a similar comparison as in Paper I, but now for 
oblique MHD shocks. 
For this we choose two representative shocks; 
namely, {\it EBJ95-D} and {\it BOEF95-1} described in \S 4.1 and \S 4.2, respectively.

In the two-fluid version of the diffusive acceleration
model the energy moment of the diffusion-convection
equation (\ref{diffcon}) is integrated from $p_2$ to $p_3$
to produce the conservation equation for CR energy; namely,
\begin{eqnarray}
{dE_c\over dt} = - \gamma_c E_c (\vec\nabla\cdot \vec u)
+ \vec\nabla\cdot (\langle \kappa \rangle \vec\nabla E_c 
- \vec u_w \gamma_c E_c)\,\\
\quad + \vec u_w\cdot\vec\nabla P_c + S_{tf},\nonumber
\label{twofld}
\end{eqnarray}
where $E_c$ is the CR energy density. 
No new approximations are
introduced in deriving equation \ref{twofld} from equation \ref{diffcon}.
It does contain three closure parameters, $\gamma_c$, $S_{tf}$ and 
$\langle\kappa\rangle$ 
that are really properties of the solution, but in practice must be
estimated {\it a priori}.
For these particular simulations it is sufficient to set 
the CR adiabatic index, $\gamma_c = \frac{5}{3}$, 
since the particle populations are entirely nonrelativistic.
The injection energy rate, $S_{tf}$ 
represents energy exchange with the thermal plasma 
(see also equation [2.5] in Paper I). 
In the diffusion-convection simulations, 
we calculate numerically an analogous injection energy rate, $S$, 
from the particle flux crossing the
low momentum boundary of the CR population at $p_2$, 
and subtract it from the thermal energy.
Then the spatially integrated injection rate for the two-fluid model, 
$I=\int S dx$, can be parameterized by 
a dimensionless two-fluid ``injection parameter'', $\eta$, 
through the relation $I = (1/2) v^2_2 \rho_1 u_1~ \eta$
where $v_2 = p_2 c$ (see equation [2.4] in Paper I).
Thus we calculate $\eta$ rather than $I$ itself
as a function of time for each shock modeled using the results of
the kinetic equation calculations.
In practice the value of $\eta$ is fairly constant over time in the 
cases we have considered,
so very comparable two-fluid solutions would be found by assuming a 
single value in each test; namely $\eta \approx 0.006$ for {\it EBJ95-D} 
and $\eta \approx 0.002$ for {\it BOEF95-1}.

To model the mean diffusion coefficient, $\langle\kappa\rangle$ (see
equation [2.13] in Paper I), 
we used the form for $\kappa(p)$ discussed in \S 3 and a simple
power-law model for the CR distribution function; namely, $f(p) \propto p^{-q}$,
where $q = 3r/(r-1)$ is the standard test-particle index expected
for diffusive acceleration and $r$ is the initial compression ratio for
the shock (see Figs. 2, 6). 
We supposed that the CR distribution extended between
$p_2$, as defined for the full diffusion-convection simulation and $p_3$ found
from the usual relation between particle energy and mean
acceleration time (\cite{lagces83};  EBJ95). In the present context
that leads to $p_3 = p_2 (1~+~t/\tau)^{1/2}$, 
where $\tau = \frac{3}{2} \kappa(p_2)/(u_1 \Delta u)$,
and $\Delta u = u_2 - u_1$, for the initial shock. 
In practice we obtained somewhat
better matches with the diffusion-convection runs by replacing the 
factor $\frac{3}{2}$
by the factor 2; that is, the distribution begins to cut off
a little below $p_3$. Our results presented here use that latter value.

Two-fluid models are intended only for dynamical studies, so the appropriate tests
are comparisons of shock structures computed by the two-fluid model and structures
computed by diffusion-convection methods (or actual shocks, if available).
The two-fluid shock structure evolution for {\it EBJ95-D}
is shown by the dotted lines in Fig. 2. 
The agreement with the diffusion-convection
simulation is good, reinforcing our earlier conclusions about the basic
validity of the two-fluid model. 
At the end of this simulation ($t = 1.2\times 10^8$ sec) $P_c$ is definitely
producing important modifications to the flow structure. 
It is still increasing, so
that a more major dynamical influence could be expected at later times.
In fact, the time asymptotic two-fluid solution for this shock should be completely
CR dominated, as can be demonstrated from comparable shock solutions in the Figure 3 
presented by Kang \& Jones (1990), or
Figure 3 of Frank, Jones \& Ryu (1995), for example. 
It is simple to demonstrate that the time-asymptotic two-fluid jump conditions
(hence, $P_c$ downstream), are independent of the value or spatial structure
of the diffusion coefficient.
For this shock, however, the time required to approach that solution from 
the initial conditions we used would be extremely long; so long, in fact, 
that the practical significance of the time-asymptotic solution is doubtful.

By contrast, the shock structure at intermediate times is influenced 
sensitively by the early time evolution of $\langle\kappa\rangle$. That grows quickly
and asymptotes to $\langle\kappa\rangle \propto (t/\tau)^2$.
The rate of dynamical shock evolution generally scales inversely as the
``diffusion time'', $t_d = \langle\kappa\rangle/u_1^2$.
That means at late times the shock structure evolves very slowly.
At intermediate times, $P_c$ is largely controlled
by the very early history of the shock; 
particularly $\langle\kappa\rangle$ and $\eta$.
So, the match we see in Fig. 2
between the diffusion-convection and two-fluid simulations is sensitive to 
the values of $\tau$ and $\kappa(p_2)$. 
That the appropriate value of $\tau$ is reasonably well predicted by the simple
test particle theory for the diffusion coefficient is an encouraging outcome. 
The minor differences between the diffusion-convection and two-fluid runs 
come, in fact, from small differences in the upstream spatial variations of 
$\langle\kappa\rangle$ modeled in the two-fluid calculations
and as computed directly from the diffusion-convection simulation. 
In the diffusion-convection simulation, 
the particle distribution tends to ``harden'' significantly upstream of 
the shock (see, \eg Kang \& Jones 1991; Paper I), so
the actual $\langle\kappa\rangle$ increases away from the shock, reducing
the value of $P_c$ as a result. 

The {\it BOEF95-1} shock has a dynamically very significant magnetic field, 
so it presents an important comparison case for two-fluid models in the 
MHD shock context.
Both the two-fluid and the diffusion-convection shock structures for 
{\it BOEF95-1} are shown in Fig. 6.
The two-fluid parameters were determined in exactly the same way as for
{\it EBJ95-D}. 
Again the agreement between the two models is very good. 
Recall that we already made a comparison between the particle velocity 
distributions from the diffusion-convection solution and 
the Ulysses observations, but that we have no detailed information about
the physical, interplanetary shock structure. 
In this case the value of $P_c$ at the shock is less than 10\% of the gas
pressure, $P_g$, by the end of the simulations ($t = 6$ minutes), so there 
is only minor modification of the shock structure by the CRs. 
In {\it BOEF95-1}, as in {\it EBJ95-D} the structure at intermediate times is 
primarily determined by the evolution of $\langle\kappa\rangle$
at the early times in the simulation. 
Again, because $\langle\kappa\rangle$
increases rapidly with time, the $t_{d}$ that determines the
rate of shock evolution becomes very long.
A direct consequence of this is that, despite the apparently
steady shock by the end of the simulation (and also the diffusion-convection
simulation), 
we are {\it not} seeing the solution that would
be determined from the steady-state two-fluid model. 

To illustrate the point, we can take advantage of the argument made earlier that the
steady-state jump conditions can be found using any convenient diffusion 
coefficient.
For this we have recomputed {\it BOEF95-1} using a constant $\langle\kappa\rangle = 0.4$,
and allowed it to evolve to a steady state. In this case we use
the criterion for a time-asymptotic solution that not only does the peak 
value of $P_c$ become steady,
but also that $P_c$ be uniform directly behind the shock.
The evolution of this shock ({\it BOEF95-1C}) and its final structure are 
shown in Fig. 7.
Early on, the evolution of {\it BOEF95-1C} and {\it BOEF95-1} are similar, 
because $\langle\kappa\rangle$ are comparable. 
However, as the {\it BOEF95-1} diffusion coefficient increases
dynamical evolution ``stalls'', while the constant diffusion coefficient
of {\it BOEF95-1C} allows it to continue directly towards the formal steady 
state solution.
The final acceleration efficiency of this shock as measured by $P_c$
is more than an order of magnitude greater than for the {\it BOEF95-1} 
simulations presented earlier or the physical shock.
Thus, it would clearly be inappropriate to apply the steady state
two-fluid model to estimate the efficiency expected in this shock for 
even moderately long times.

The discrepancy is not an indication of basic flaws in the two-fluid model, 
but rather that time-asymptotic solutions are not very relevant in this 
situation. 
The key question becomes how to choose, in the two-fluid model, 
a meaningful set of closure parameters for a time dependent calculation. 
In both of the tests conducted here it seems adequate to
apply simple test-particle models to those parameters, 
because the shock structure is not sufficiently modified on short times. 
Over longer times that convenience may become dubious, but, since the shock 
properties at moderately late times are largely set by
the conditions early in the shock evolution, 
this breakdown may not be crucial in many instances. 
It is important to know, then, if the shock structure should
evolve quickly on time scales of interest. 
That is something we can hope to estimate reasonably well using the 
standard test-particle approach. These issues
are more important if we expect a steep momentum dependence to the diffusion 
coefficient, as in the cases studied here, so a more basic understanding of the
evolution of the resonant Alfv\'en wave field becomes important, as well.

\section{Conclusion}

In order to study the particle acceleration in oblique 
magnetohydrodynamic (MHD) shocks,  
we have implemented the existing diffusion-convection methods 
of Kang \& Jones 1991 into a full MHD code,
and adopted a ``thermal leakage'' type injection model 
introduced by Kang \& Jones 1995 (Paper I).
In our injection model, the distribution of the suprathermal 
particles which cannot be treated properly with 
the diffusion-convection method was assumed to match smoothly
onto the Maxwellian distribution of the gas particles.
The matching condition is controlled by a free parameter $c_1$,
which in turn determines the particle injection rate into the 
CR population.
Firstly, we have calculated the MHD shocks for various field obliquities 
considered by Ellison, Baring \& Jones (1995), in order to study the dependence of
the injection efficiency on some shock properties via test-particle
Monte Carlo simulations.
By adjusting the free parameter $c_1$ of our injection model over a modest range
we were able to demonstrate that our numerical technique  
can, in fact, produce particle spectra comparable to theirs.
Secondly, we have reproduced the proton flux distributions
at oblique interplanetary shocks observed {\it in situ} by the Ulysses spacecraft. 
These shocks have also been simulated previously by Baring \etal 1995,
via test-particle Monte Carlo technique. 
We adopted the shock parameters chosen by them to match direct observations. 
To obtain good agreement with the observations, the numerical shock
thickness for these quasi-perpendicular shocks has to be about (1-2) times 
the thermal gyro-radius.
This is consistent with the conclusion of Baring \etal that 
strong scattering turbulence was present in these interplanetary shocks.

The Monte Carlo technique treats both thermal and cosmic ray particles
by the same scattering law, so the injection process comes about
naturally via the acceleration of thermal particles to higher energies.
In contrast, injection cannot be determined self-consistently 
through diffusion-convection models for cosmic-ray transport, since particles
at momenta where injection occurs do not form an isotropic distribution
and the diffusion approximation is not valid.
Our model works around this by introducing a free parameter that establishes the
momentum at which the suprathermal distribution must match onto the thermal 
distribution behind the shock.
One might be concerned that the particle distribution at intermediate energies
(between the thermal peak energy and energies much greater than thermal
peak energy) would be dependent upon the details of the injection 
process.
Our simulations, however, seem to indicate that the dependence is not 
sensitive enough to make a clear distinction between the 
particle spectra simulated with our diffusive injection model 
and existing observations, or the particle spectra simulated with
Monte Carlo techniques.
This leads us to the tentative, but encouraging conclusion that
a simple, macro-physical model like ours can offer a practical
compromise between {\it ad hoc} injection and injection models built
to include microphysical details.

Comparison tests presented here and in Paper I have shown that,
in our model, 
the injection process and its rate are mostly determined by the 
numerical shock thickness in terms of the thermal gyro-radius 
and the momentum, $p_1=c_1~p_{th}$,
where the suprathermal particle distribution matches onto the Maxwellian
distribution.  Presumably both of these are affected in a detailed
model by the strength of Alfv\'en turbulence (\eg Malkov \& V\"olk 1995).
Under the assumption that the shock thickness is about the mean
scattering length of the thermal peak momentum, $p_{th}$, for
quasi-parallel shocks and the thermal gyro radius, $r_g(p_{th})$ for
quasi-perpendicular shocks,
the appropriate value of $c_1$ lies between 1.5-2.
A smaller $c_1$ leads to larger injection rates, because it allows a larger ``injection
pool'' of diffusive particles to form.
EBJ95 found in Monte Carlo simulations that the rate of particle injection 
decreased strongly with increasing obliquity for strong shocks, 
unless the cross-field diffusion was strong. In our model that
is effected by increasing the parameter $c_1$ with obliquity  or
decreasing it towards smaller Mach number or stronger Alfv\'en
turbulence.
More quantitative prediction requires much further work, however.
We note also that the injection efficiencies given in EBJ95, which
were based on test-particle
simulations, are overestimated compared to dynamical
shocks, especially for strong, quasi-parallel shocks,
since the CR energy density predicted by those simulations is significant
enough that the CRs would modify the shock dynamics. 
In particular, the thermal particle
distribution would be colder and the subshock velocity jump would
be smaller in self-consistent dynamic calculations. 
These differences between test-particle and dynamical shocks were 
confirmed by the recent Monte Carlo simulations of
EBJ96. 

In Paper I,
we showed that diffusive acceleration numerical methods applied to parallel shocks
produce similar shock structures and particle distributions
compared to Monte Carlo and hybrid plasma methods, and that they matched direct
observations at the earth's bow shock. 
The comparisons reported here for oblique shocks strengthen the
important conclusions of Paper I 
that the essential physics of the particle injection and 
acceleration can be captured by each of these diverse 
computational methods, and that they are all practical
and complementary tools for understanding the physics of
diffusive shock acceleration.
This also implies that our numerical approach provides a way to 
incorporate naturally the injection process into the existing 
diffusion-convection technique.
The advantages of this formalism distinguishing it from Monte
Carlo or plasma methods are that it is time-dependent, in addition to being
a fully dynamically {self-consistent}
MHD diffusion-convection technique, so that
it can be used for evolving and structurally complex flows.
In upcoming papers, we will use it 
to study the acceleration efficiency
and the nonlinear back reaction of CR pressure on the shock dynamics
in various astrophysical shock waves.
  
We also simulated a pair of oblique two-fluid shocks. 
Each was constructed exactly as for one of the diffusion-convection simulations
reported, with the required closure parameters determined from simple test-particle
considerations. The dynamical properties of the two-fluid shocks are quite
consistent with the diffusion-convection solutions. 
These simulations demonstrate in 
the oblique shock context the basic validity of the two-fluid method. 
We emphasize, however, that {\it steady state} two-fluid solutions 
may not be applicable, even when the shock structures appear to be steady. 
If the cosmic-ray diffusion coefficient has
a strong momentum dependence, the rate of shock evolution can become 
very slow, so that while a shock may appear dynamically steady,
in practical terms the time-asymptotic solution is not likely to be reached 
for a long time. Then the dynamical conditions creating the shock may 
very likely have changed, requiring the shock to readjust once more.

\acknowledgments 

We are grateful to Matthew Baring for illuminating discussions about oblique CR
shocks and Monte Carlo methods and for useful suggestions that helped us 
improve the manuscript. 
This work was supported in part at the University of Minnesota by the 
NSF through grant AST-9318959, by NASA through grants NAGW-2548 and
NAG5-50505 and by the 
University of Minnesota Supercomputer Institute.  
HK is supported in part at Pusan National University 
by the Korean Research Foundation through the Brain Pool Program.
                    
\vfill\break

\clearpage          
\begin{figure*}      
\epsfysize=7.0in\epsfbox[54 72 560 750]{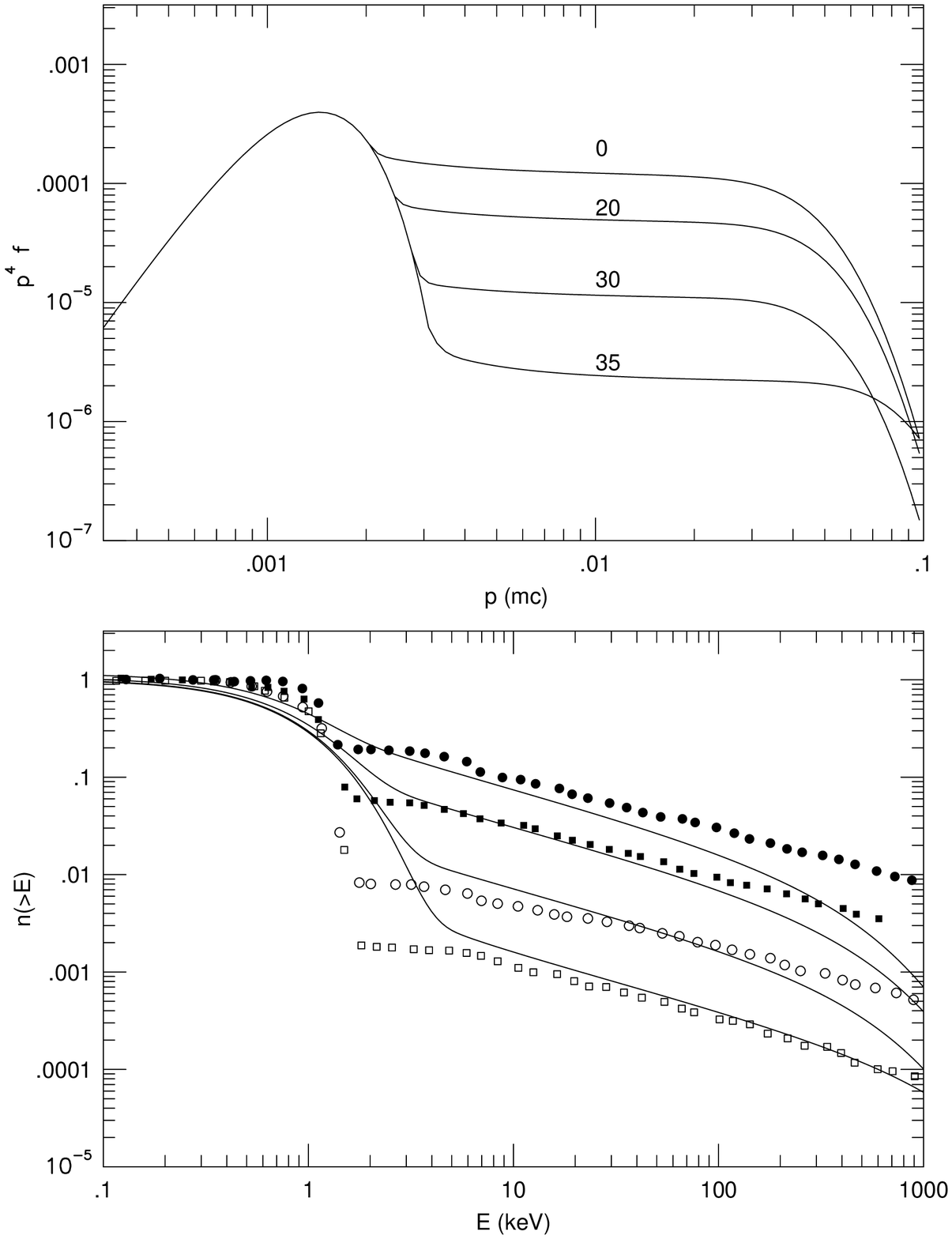}
\caption{
Top panel shows the distribution function $g=f(x_s,p)p^4$ versus the
particle momentum $p$ in units of $mc$ for particles
at the shock for {\it EBJ95} test runs. 
The lines are labeled by the value of the upstream obliquity angle $\theta_1$.
The results are shown at $(t/10^8 s) = 12,8,6,$ and 4 for $\theta_1=$ 
0\arcdeg, 20\arcdeg, 30\arcdeg and 35\arcdeg, respectively.
See text for the shock parameters. 
Bottom panel shows the integral density distributions calculated
from the momentum distribution function shown in the top panel.
The open and filled circles and squares are representative points of
EBJ95 Monte Carlo simulation results of the same shock conditions (from their Fig. 5). 
}
\end{figure*}        
                    
\begin{figure*}      
\epsfysize=7.0in\epsfbox[54 72 560 750]{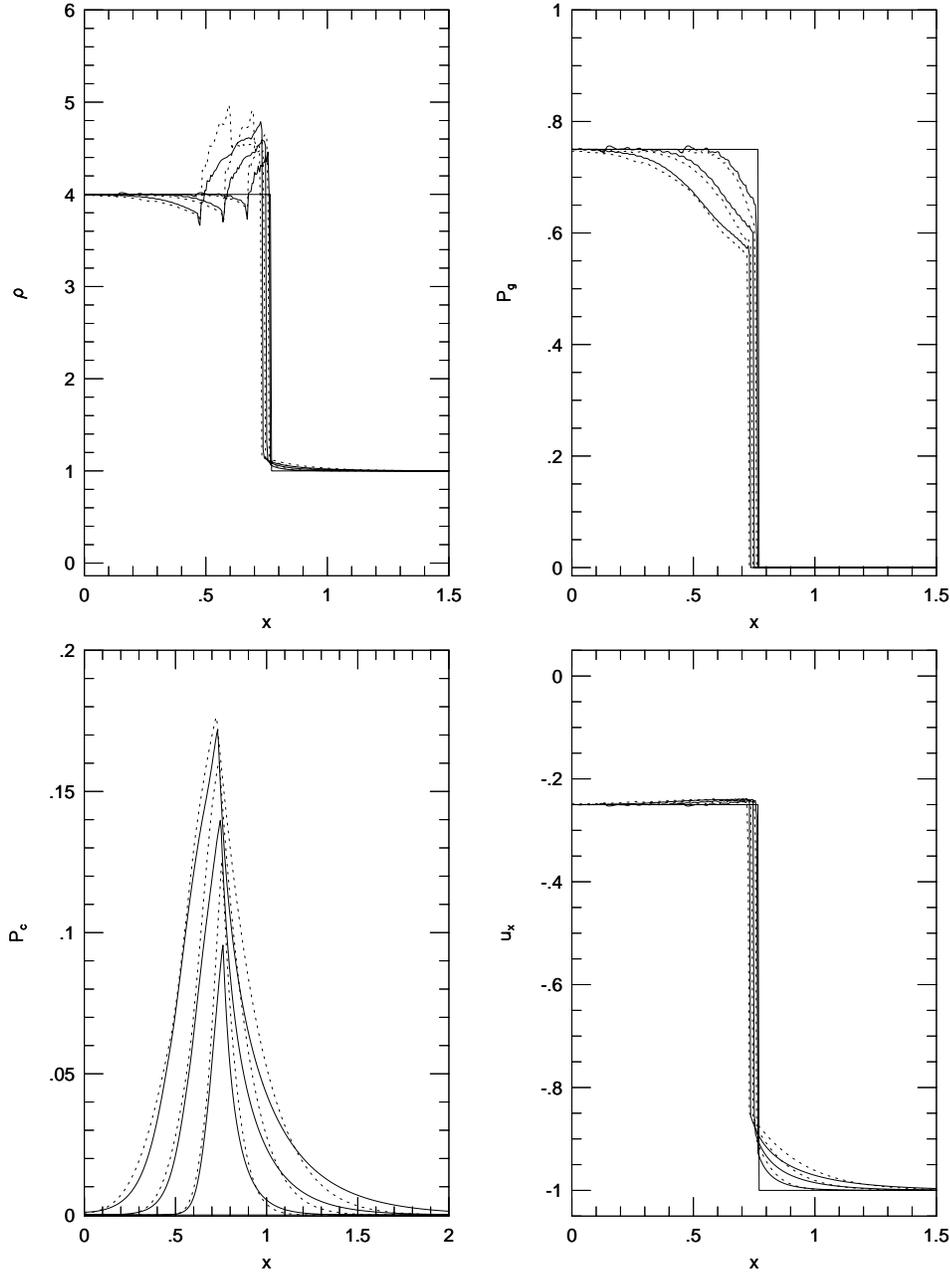}
\caption{
The shock flow structure of the model {\it EBJ95-D} shock at $(t/10^8 s)=0.4,
0.8, 1.2$  in
our fully dynamical simulations (solid line).
The initial condition is specified by the pure MHD shock jump.
The two-fluid solution for the same shock as discussed in \S4.3
is shown by the dotted line. This is a Mach 100 shock, with a very
weak magnetic field at an upstream obliquity, $\theta_1 = 30$\arcdeg.
}
\end{figure*}

\begin{figure*}      
\epsfysize=7.0in\epsfbox[54 72 560 750]{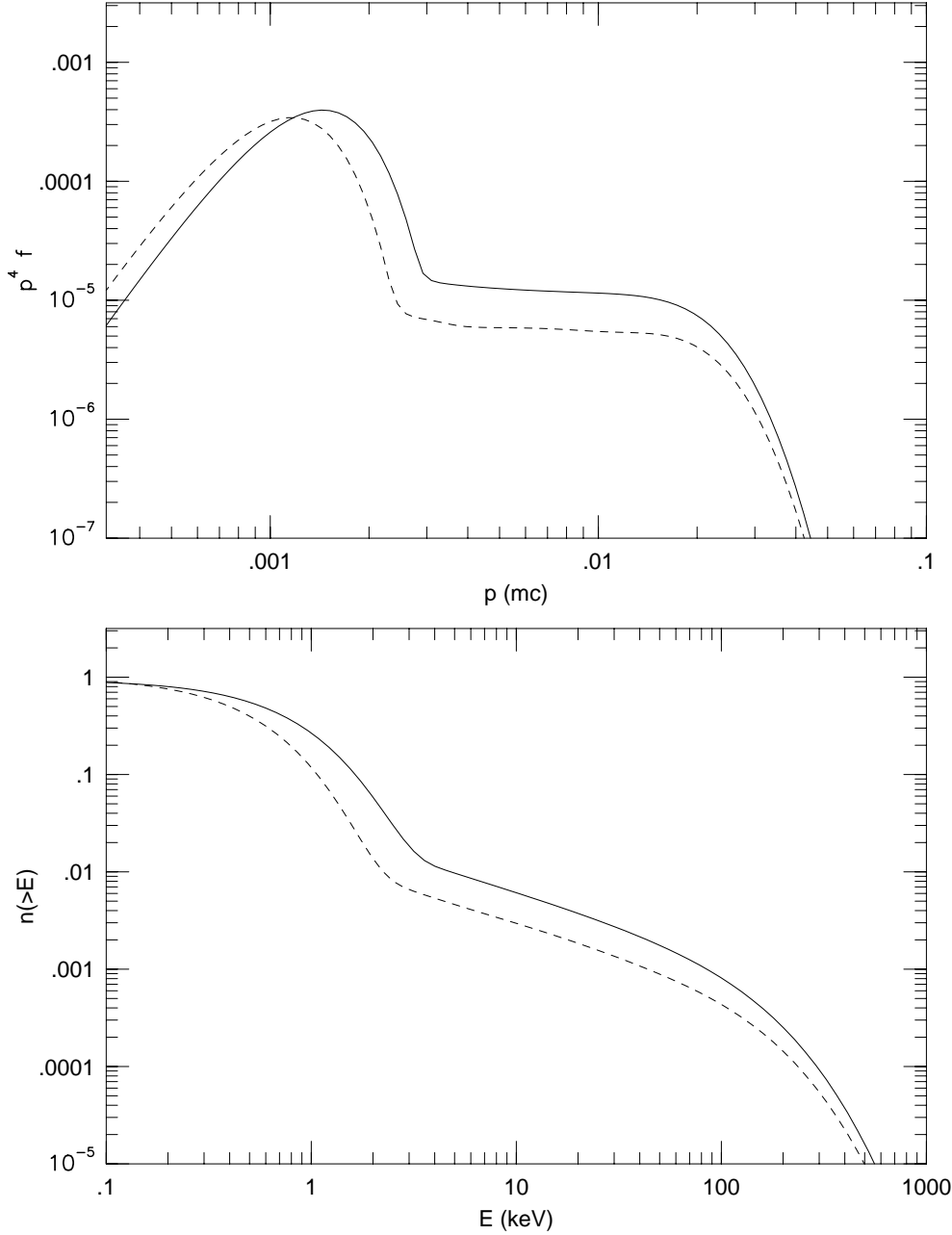}
\caption{
Top panel shows the distribution function $g=f(x_s,p)p^4$ versus the
particle momentum $p$ 
at the shock for {\it EBJ95} test runs at $(t/10^8 s)=1.2$.
The obliquity, $\theta_1=30\arcdeg$ and the injection parameter, $c_1=2.0$
The solid line is the spectrum from the test-particle run, while
the dotted of line is for fully dynamic runs.
Bottom panel shows the integral density distributions calculated
from the distribution functions shown in the top panel.
}
\end{figure*}  
      
\begin{figure*}      
\epsfysize=7.0in\epsfbox[54 72 560 750]{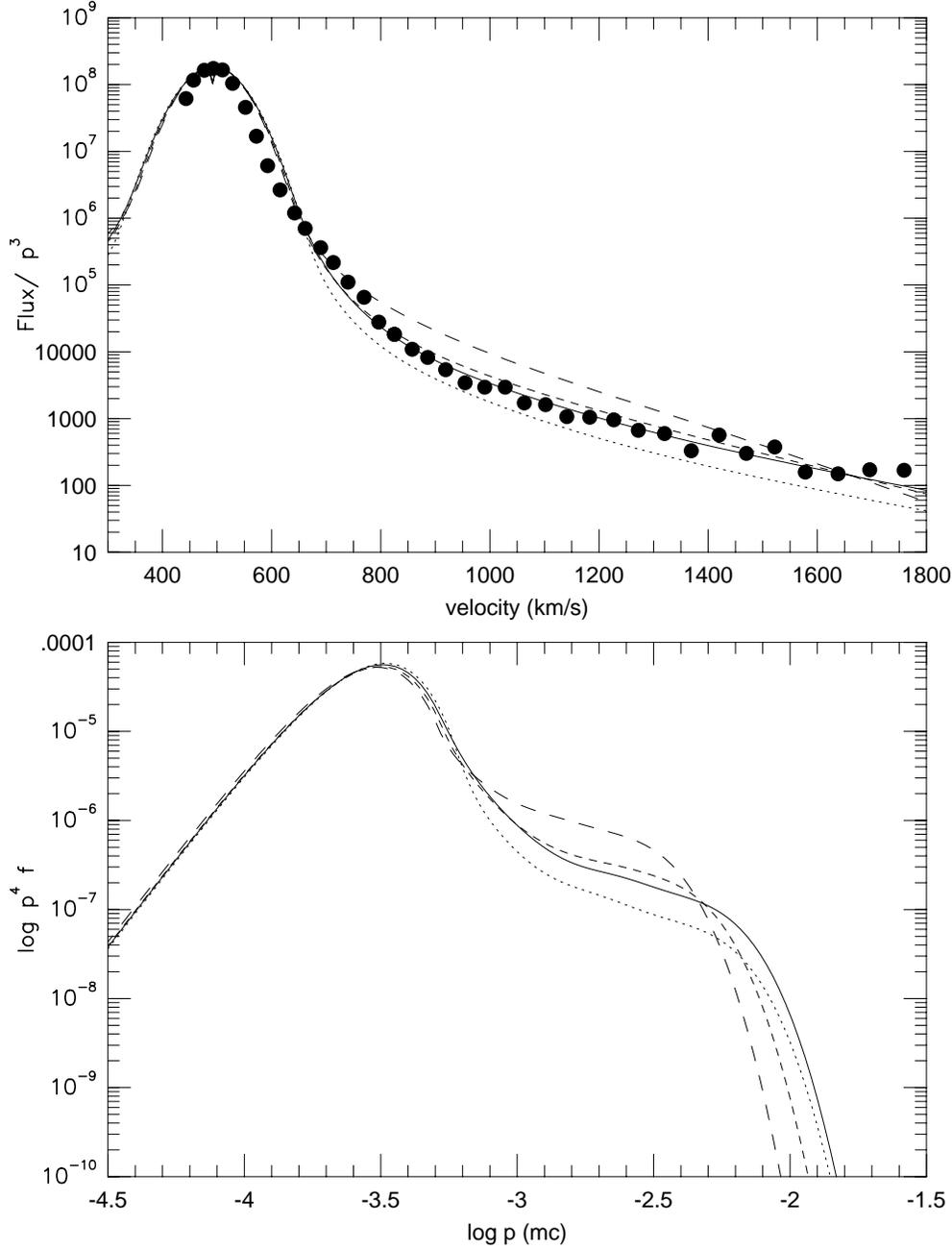}
\caption{ 
Simulated omni-directional particle flux in the Ulysses spacecraft frame divided by
the particle momentum cubed, $p^3$ (upper), and the particle distribution
function $f(p)$ (lower) for {\it BOEF95-1} shock at $t=6$ minutes.
The solid line is for $N=4$, dashed line for $N=20$, and long-dashed
line for $N=40$. The value of $c_1=1.6$ for these three runs.
The dotted line is for $N=4$ and $c_1=2.0$
}
\end{figure*}        
                    
\begin{figure*}      
\epsfysize=7.0in\epsfbox[54 72 560 750]{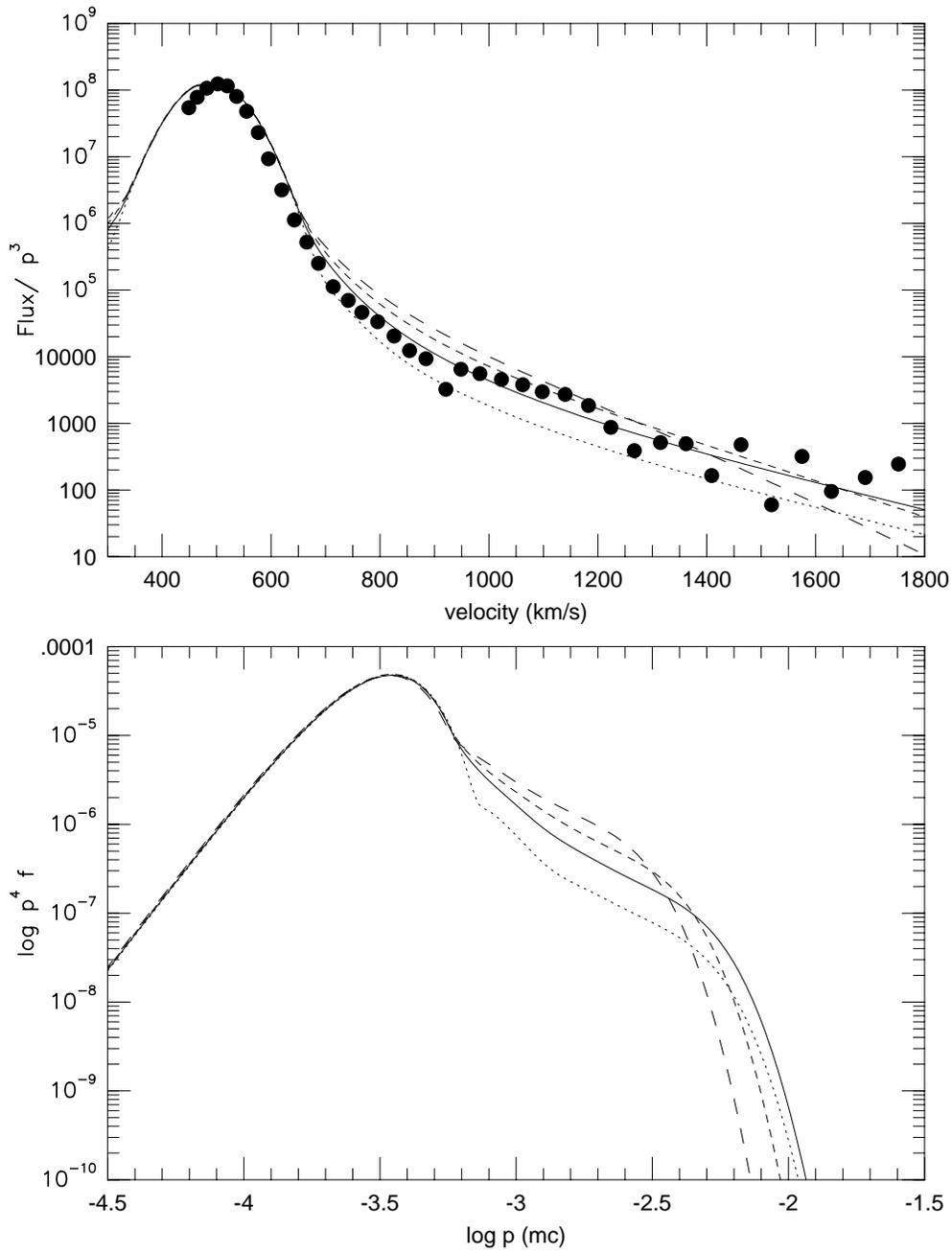}
\caption{
Same as Fig. 4, except for {\it BOEF95-2} shock at $t=10$ minutes.
The solid line is for $N=9$, dashed line for $N=20$, and long-dashed
line for $N=40$. The value of $c_1=1.6$ for these three runs.
The dotted line is for $N=9$ and $c_1=2.0$
}
\end{figure*}        

\begin{figure*}      
\epsfysize=7.0in\epsfbox[54 72 560 750]{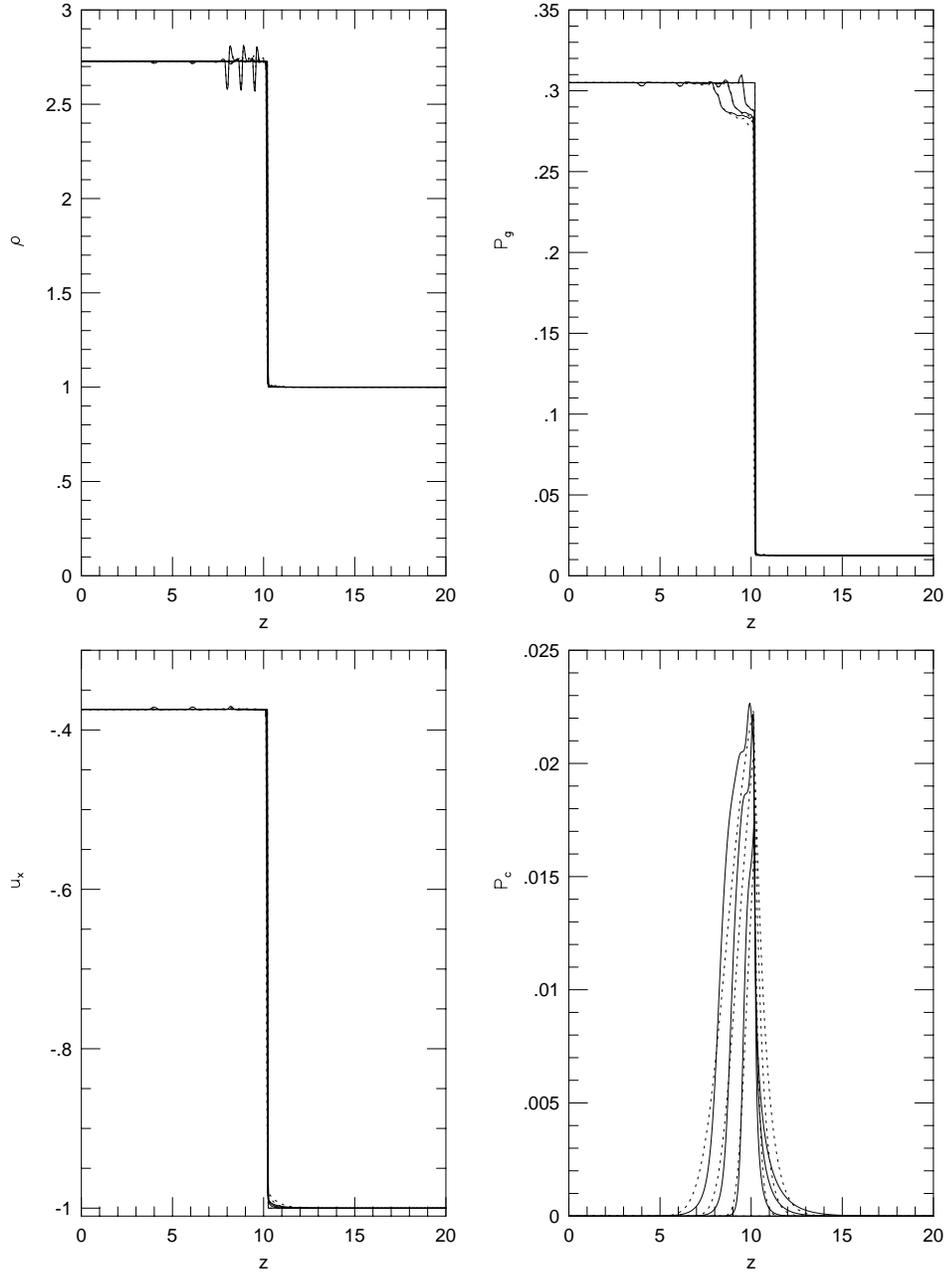}
\caption{
The shock structure computed for the {\it BOEF95-1} shock at $t = 2,~4,~6$ minutes. The
diffusion-convection solution is shown by the solid lines and the two-fluid
solution by the dotted lines. The two-fluid solution uses a mean
diffusion coefficient that evolves in time according to a simple
test particle model for the distribution function.
The shock initial conditions are indicated by the discontinuous
curves. Details are given in the text.
}
\end{figure*}    

\begin{figure*}      
\epsfysize=7.0in\epsfbox[54 72 560 750]{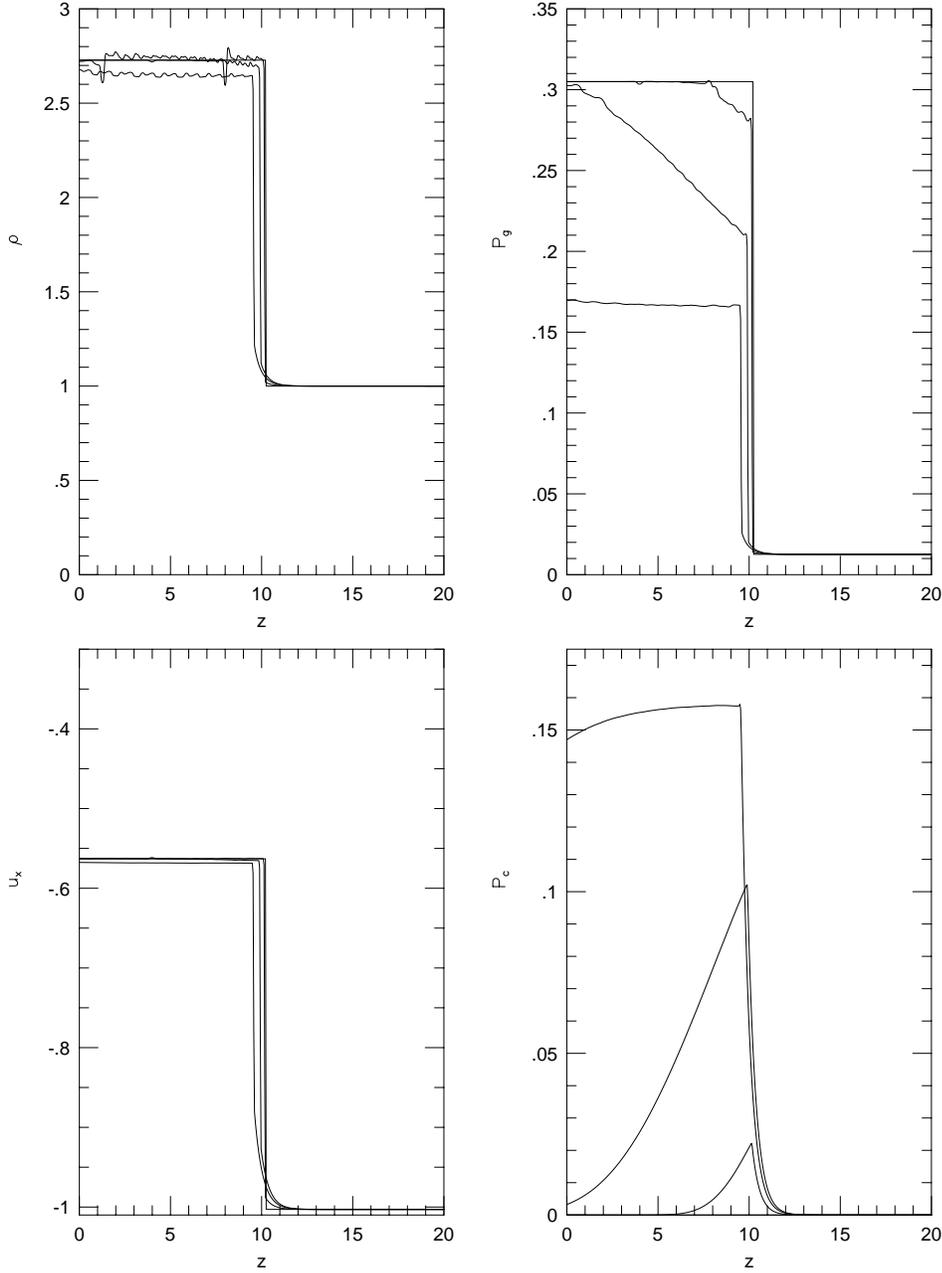}
\caption{
Evolution of the shock structure for the shock {\it BOEF95-1C}. This is a two-fluid model
shock and differs from the two-fluid shock shown in Fig. 6 only in the use here
of a constant diffusion coefficient, $\langle\kappa\rangle = 0.4$. Times represented
are $t = 0,~6,~24,~72$. For the last time the shock has approached its time-asymptotic
limit.
}
\end{figure*}

\end{document}